\documentclass[a4paper, pra, twocolumn,superscriptaddress,showpacs]{revtex4}

\pdfoutput=1
\usepackage{amssymb}
\usepackage{amsmath}
\usepackage{epsfig}
\usepackage{color}
\usepackage{graphics, graphicx}
\usepackage{bbold}
\usepackage{psfrag}
\usepackage{mathcomp}
\usepackage{subfigure}
\usepackage{verbatim}
\usepackage{color}
\usepackage[colorlinks,citecolor=blue]{hyperref}

\begin{document}

\date{\today}
\title{BCS-BEC crossover and quantum phase transition in an ultracold Fermi gas under spin-orbit coupling}
\author{Fan Wu}
\affiliation{Key Laboratory of Quantum Information, University of Science and Technology of China,
CAS, Hefei, Anhui, 230026, People's Republic of China}
\affiliation{Synergetic Innovation Center of Quantum Information and Quantum Physics, University of Science and Technology of China, Hefei, Anhui 230026, People's Republic of China}
\author{Ren Zhang}
\affiliation{Department of Physics, Renmin University of China, Beijing 100872, People's Republic of China}
\author{Tian-Shu Deng}
\affiliation{Key Laboratory of Quantum Information, University of Science and Technology of China,
CAS, Hefei, Anhui, 230026, People's Republic of China}
\affiliation{Synergetic Innovation Center of Quantum Information and Quantum Physics, University of Science and Technology of China, Hefei, Anhui 230026, People's Republic of China}
\author{Wei Zhang}
\email{wzhangl@ruc.edu.cn}
\affiliation{Department of Physics, Renmin University of China, Beijing 100872, People's Republic of China}
\affiliation{Beijing Key Laboratory of Opto-electronic Functional Materials and Micro-nano Devices,
Beijing 100872, People's Republic of China}
\author{Wei Yi}
\email{wyiz@ustc.edu.cn}
\affiliation{Key Laboratory of Quantum Information, University of Science and Technology of China,
CAS, Hefei, Anhui, 230026, People's Republic of China}
\affiliation{Synergetic Innovation Center of Quantum Information and Quantum Physics, University of Science and Technology of China, Hefei, Anhui 230026, People's Republic of China}
\author{Guang-Can Guo}
\affiliation{Key Laboratory of Quantum Information, University of Science and Technology of China,
CAS, Hefei, Anhui, 230026, People's Republic of China}
\affiliation{Synergetic Innovation Center of Quantum Information and Quantum Physics, University of Science and Technology of China, Hefei, Anhui 230026, People's Republic of China}

\begin{abstract}
In this work, we study the BCS-BEC crossover and quantum phase transition in a Fermi gas under Rashba spin-orbit coupling close to a Feshbach resonance. By adopting a two-channel model, we take into account of the closed channel molecules, and show that combined with spin-orbit coupling, a finite background scattering in the open channel can lead to two branches of solution for both the two-body and the many-body ground states. The branching of the two-body bound state solution originates from the avoided crossing between bound states in the open and the closed channels, respectively. For the many-body states, we identify a quantum phase transition in the upper branch regardless of the sign of the background scattering length, which is in clear contrast to the case without spin-orbit coupling. For systems with negative background scattering length in particular, we show that the bound state in the open channel, and hence the quantum phase transition in the upper branch, are induced by spin-orbit coupling. We then characterize the critical detuning of the quantum phase transition for both positive and negative background scattering lengths, and demonstrate the optimal parameters for the critical point to be probed experimentally.
\end{abstract}

\maketitle

\section{Introduction}

Synthetic spin-orbit coupling (SOC), a recent addition to the toolbox available for quantum simulation in ultracold atomic gases, can give rise to interesting two-body and many-body properties by modifying the single-particle dispersion spectrum of the underlying system \cite{gauge2exp,fermisocexp1,fermisocexp2,fermisocfeshnist,fermisocexpfeshbach}. In ultracold Fermi gases, it has been shown that the implementation of synthetic SOC can lead to an unconventional superfluid with non-trivial topological features, or a superfluid with non-zero center-of-mass (CoM) momentum, or a combination of both, depending on the spatial dimensions of the gas, and on the form of synthetic SOC implemented \cite{soc3,TFZheng,TFWei,Ren_W,chenggang,huhuitwo-c,WF_A,zhang,sato,chuanwei,soc4,soc6,iskin,thermo,2d2,2d1,melo,wmliu, helianyi,wy2d,xiaosen,wypolaron,iskinnistsoc,puhantwobody,shenoy,wyfflo,melosupp2,puhan3dsoc, XF,xiongjun, hufflo,chuanweifflo,bdg1,bdg2,HUHUI_2,zhangpeng,zhangpeng2,shenoy2}. Most of these studies have assumed the system to be close to a Feshbach resonance, so that the interaction is tunable via an external magnetic field.
However, to characterize the Feshbach resonance, most of the previous studies on spin-orbit coupled Fermi systems have adopted a single-channel model \cite{TFZheng,TFWei,WF_A,zhang,sato, chuanwei,soc4,soc6,iskin,thermo,2d2,2d1,melo,wmliu,helianyi,wy2d,xiaosen,wypolaron,iskinnistsoc,puhantwobody, wyfflo,melosupp2,puhan3dsoc,XF,xiongjun,hufflo,chuanweifflo,bdg1,bdg2,HUHUI_2,zhangpeng,zhangpeng2,shenoy2}. While on a phenomenological level,
a two-channel model is more appropriate, where the Feshbach resonance is described as a multi-channel resonant scattering process when the bound state in a closed channel is tuned close to the continuum threshold of an open channel \cite{FBR}. The two-channel model reduces to a single-channel model only when the population of the closed channel molecule becomes negligible, which is not always the case, particularly under SOC \cite{huhuitwo-c,Ren_W}.

Recently, there have been several studies using two-channel models for the characterization of spin-orbit coupled Fermi gases near a Feshbach resonance \cite{huhuitwo-c,chenggang,Ren_W,shenoy}. A particularly interesting finding is that the SOC can induce a new branch of two-body bound state. While it has been reported before that in the absence of SOC, two branches of bound state can be found near a Feshbach resonance for a Fermi gas with positive background scattering length, the extra bound state under negative background scattering length is purely induced by SOC. The existence of this new two-body bound state should leave signatures on the many-body level. Indeed, for a Fermi gas without SOC, a quantum phase transition exists for a positive background scattering length, which is intimately connected with the corresponding two-body bound state \cite{WY_D}. We expect that similar phase transitions may appear in a spin-orbit coupled Fermi gas when the proper two-channel resonant scattering process is considered.

In this work, we study an ultracold Fermi gas under Rashba SOC close to a Feshbach resonance using a two-channel model. We first confirm the two-body calculations in Refs. \cite{shenoy,chenggang}, and study the branching of the two-body bound state in the presence of a finite background scattering length. For a positive background scattering length, the open channel supports a bound state even in the absence of SOC, and the two branches of two-body bound state originate from the avoided level crossing between the bound states in the open and the closed channels. For a negative background scattering length, an SOC-induced bound state emerges in the open channel for any finite SOC. The SOC-induced bound state then couples with the bound state in the closed channel, also leading to two branches of bound state.

With these understandings, we characterize many-body properties of the system using a Bardeen-Cooper-Schrieffer (BCS) mean field approach. As expected, we find two-branches of many-body solutions, the upper and the lower branch, for any finite background scattering length. With a positive background scattering length, we find that the lower branch is always bosonic with negative chemical potential, essentially a condensate of tightly bound molecules. A quantum phase transition exists in the upper branch, across which the ground state of the Fermi gas changes from a superfluid state to a normal state. The position of the phase transition can be controlled by tuning the SOC strength. We also notice that by tuning the interaction or the SOC strength, a Bardeen-Cooper-Schrieffer to Bose-Einstein condensate (BCS-BEC) crossover occurs in the upper branch.
With a negative background scattering length, the upper branch emerges from the scattering threshold on the low-field-side of the Feshbach resonance via a quantum phase transition for any finite SOC.

We discuss in detail the many-body properties of different branches under various parameters, and characterize the critical detuning for the onset of the quantum phase transition. We show that the results in this work, the quantum phase transitions in particular, should best be observed in narrow Feshbach resonances under appropriate SOC. While experimentally, only an equal mixture of Rashba and Dresselhaus SOC has been realized in cold atomic gases \cite{gauge2exp,fermisocexp1,fermisocexp2}, there have been various proposals for realizing the Rashba-type SOC \cite{xiongjun,rashbagen1,rashbagen2,rashbagen3}. With the recent experimental implementation of Feshbach resonance in spin-orbit coupled degenerate Fermi gases \cite{fermisocfeshnist,fermisocexpfeshbach}, we expect that the SOC-induced quantum phase transition reported here can be experimentally probed in the future.

The paper is organized as the following: in Sec. II, we introduce the two-channel model for a Fermi gas under Rashba SOC and close to a Feshbach resonance. In Sec. III, we study the two-body bound state solutions under a finite background scattering length. In Sec. IV, we discuss in detail the many-body ground state of the two-channel model using the standard BCS mean-field theory. For a finite background scattering length and a finite SOC, there are typically two branches of ground state,
where a quantum phase transition can be identified in the upper branch. We then characterize the critical point of the quantum phase transition in Sec. V, and finally summarize in Sec. VI.


\section{Two-channel model}
We consider a three-dimensional two-component Fermi gas close to a Feshbach resonance under Rashba SOC. This system can be described by a two-channel model
\begin{align}
H=H_0+H_{\rm SOC}+H_{\rm bf}+H_{\rm int},\label{eqn:2-chanH}
\end{align}
where the terms take the following forms
\begin{align}
H_0=&\sum_{\mathbf{k},\sigma=\uparrow,\downarrow}\epsilon_{\mathbf{k}}a^{\dag}_{\mathbf{k}\sigma}a_{\mathbf{k}\sigma}+ \sum_{\mathbf{q}}({\gamma}+\frac{\epsilon_{\mathbf{q}}}{2})b^{\dag}_{\mathbf{q}}b_{\mathbf{q}},\nonumber \\
H_{\rm SOC}=&\sum_{\mathbf{k}}{\alpha}[(k_x-ik_y)a^{\dag}_{\mathbf{k},\uparrow}a_{\mathbf{k},\downarrow}+h.c.],\nonumber \\
H_{\rm bf}=&\frac{{g}}{\sqrt{V}}\sum_{\mathbf{k},\mathbf{q}}(a^{\dag}_{\mathbf{k}+\mathbf{q}\uparrow} a^{\dag}_{-\mathbf{k}+\mathbf{q}\downarrow}b_{\mathbf{q}}+h.c.),
\nonumber \\
H_{\rm int}=&\frac{{U}}{V}\sum_{\mathbf{k},\mathbf{k}',\mathbf{q}}a^{\dag}_{\mathbf{k}+\mathbf{q}\uparrow} a^{\dag}_{-\mathbf{k}+\mathbf{q}\downarrow}a_{-\mathbf{k}'+\mathbf{q}\downarrow}a_{\mathbf{k}'+\mathbf{q}\uparrow}.
\nonumber
\end{align}
Here, $a_{\mathbf{k},\sigma}$($a^{\dag}_{\mathbf{k},\sigma}$) is the annihilation (creation) operator for atoms with pseudo-spin $\sigma$ and momentum ${\bf k}$,
$\epsilon_{\mathbf{k}}=\hbar^2k^2/2m$ is the single fermion dispersion, $b_{\mathbf{q}}$($b^{\dag}_{\mathbf{q}}$) is the annihilation (creation) operator for the closed channel molecules, ${\alpha}$ is the Rashba SOC strength, $V$ is the quantization volume, and $h.c.$ stands for Hermitian conjugate.
The bare atom-molecule coupling rate ${g}$, the bare background interaction rate ${U}$, and the bare detuning ${\gamma}$ are connected with the physical ones $\{{g}_p, {U}_p, {\gamma}_p\}$ through the standard renormalization relations: ${U}=\Gamma {U}_p, {g}=\Gamma {g}_p, {\gamma}={\gamma}_p-\Gamma {g}^2_p/U_c$, where $\Gamma=(1+{U}_p/U_c)^{-1}, U^{-1}_c=-\sum_{\mathbf{k}}1/2\epsilon_{\mathbf{k}}$, and ${U}_p=4\pi\hbar^2a_{\rm bg}/m, {g}^2_p=4\pi\hbar^2a_{\rm bg} W\mu_{\rm co}/m$, and ${\gamma}_p=\mu_{\rm co}(B-B_0)$ \cite{crossoverreview}. Here, $a_{\rm bg}$ is the background scattering length in the open channel, $W$ is the Feshbach resonance width, $\mu_{\rm co}$ is the magnetic moment difference between the closed and open channels, and $B-B_0$ is the magnetic field detuning with $B_0$ the Feshbach resonance point.

To be consistent with the experimental parameters, we adopt the unit of energy as $E_0$, and define the unit of the momentum $k_0$ and the unit of density $n_0$ as
\begin{equation}
k_0=\sqrt{\frac{2mE_0}{\hbar^2 }},\quad\quad n_0=\frac{k^3_0}{3\pi^2}.
\end{equation}
We then obtain a dimensionless version of the parameters in the Hamiltonian, which will be used in the following discussion.


\section{Two-body bound states}

In this section, we investigate the two-body bound-state solution under the Hamiltonian Eq. (\ref{eqn:2-chanH}).
Due to the presence of SOC, the relative motion of the fermions is dependent on the CoM motion. As a result,
the bound-state energy also acquires dependence on the CoM momentum. For the lowest energy case of zero
CoM momentum, the two-body bound state wave function can be written as
\begin{eqnarray}
|\Psi\rangle &=& \bigg\{\beta b_0^{\dagger}+\sum_{{\bf k}}{}^{\prime}\Big[\eta^{\uparrow\downarrow}({\bf k})a_{{\bf k},\uparrow}^{\dagger}
a_{-{\bf k},\downarrow}^{\dagger}+\eta^{\downarrow\uparrow}({\bf k})a_{{\bf k},\downarrow}^{\dagger}
a_{-{\bf k},\uparrow}^{\dagger} \nonumber\\
&& +\eta^{\uparrow\uparrow}({\bf k})a_{{\bf k},\uparrow}^{\dagger}
a_{-{\bf k},\uparrow}^{\dagger}+\eta^{\downarrow\downarrow}({\bf k})a_{{\bf k},\downarrow}^{\dagger}
a_{-{\bf k},\downarrow}^{\dagger}\Big] \bigg\} |0\rangle,
\end{eqnarray}
where $\beta$ and $\eta^{\sigma\sigma'}$ denote the closed channel and open channel coefficients, respectively,
and the summation over momentum space $\sum_{\bf k}^\prime$ runs over half of the momentum space with
$k_y>0$.

By solving the Shr\"{o}dinger's equation $H|\Psi\rangle=E|\Psi\rangle$ and matching coefficients, we obtain the equation
for the two-body binding energy $E$
\begin{align}
\left[{U}_p-\frac{{g}_p^2}{{\gamma}_p-E}\right]^{-1}={\cal S}({\alpha},E),
\end{align}
where ${\cal S}({\alpha},E)$ is defined as,
\begin{eqnarray}
{\cal S}({\alpha},E)\equiv \frac{3 \pi }{8 \sqrt{2}}
\left[\sqrt{-E}-\frac{\alpha}{\sqrt{2}}{\rm arctanh}\left(\frac{{\alpha}}{\sqrt{-2 E}}\right)\right].
\end{eqnarray}

\begin{figure}
\centering
\includegraphics[width=8.4cm]{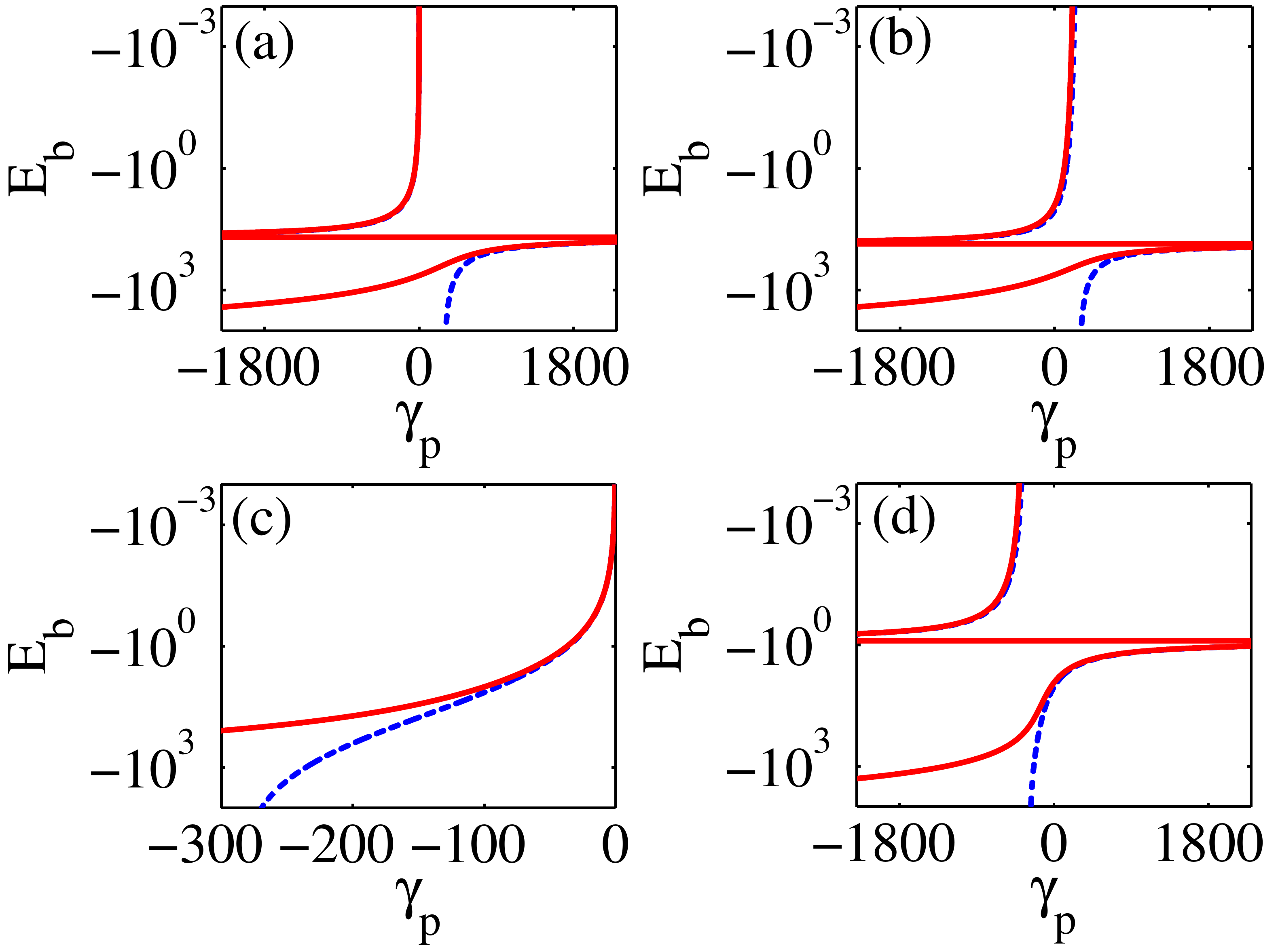}
\caption{(Color online) Two-body binding energy as a function of detuning for (a-b) positive background scattering length,
and (c-d) negative background scattering length. The results for a two-channel model (solid, red) are
compared with those from a single-channel model (dashed, blue). Using the unit system defined within the text,
the dimensionless atom-molecule coupling constant $g_p=7$ for all panels. Other parameters
are (a) $U_p=0.17,\alpha=0$; (b) $U_p=0.17,\alpha=5$; (c) $U_p=-0.17,\alpha=0$; (d) $U_p=-0.17,\alpha=5$.}
\label{fig:rashba}
\end{figure}

In Fig.~\ref{fig:rashba}, we plot typical results for the two-body binding energy $E_b \equiv E - E_{\rm th}$
for both cases of positive and negative background scattering lengths, where $E_{\rm th}=-\alpha^2/2$ is the threshold.
In the absence of SOC [see Fig.~\ref{fig:rashba}(a) and (c)], there are two branches of bound state solution for
positive background scattering length, while there is only one branch for negative background scattering length.
This is consistent with the calculations of Ref. \cite{WY_D}. For finite SOC, an additional branch of bound state emerges
for the case with negative background scattering length [see Fig.~\ref{fig:rashba}(d)], which is consistent with
the results of Ref. \cite{shenoy,chenggang}. As a comparison, results obtained from a single-channel model
with corresponding parameters are also shown.

The position of the bound state threshold in the upper branch can be determined analytically for both cases of
positive and negative background scattering length, leading to
\begin{eqnarray}
\gamma_c=-\frac{\alpha^2}{2}+\frac{g_p^{2}}{U_p}.
\end{eqnarray}
It is clear that the positions of the bound state threshold in both cases are pushed towards the BEC-side of the Feshbach resonance as the SOC strength increases.
This is the direct result of decreasing threshold energy with increasing SOC.

In the large-detuning limit, the binding energy in either branch asymptotically approaches a common value $E_{\inf}$, which
is determined by the following equation,
\begin{eqnarray}
{U}_p=\frac{16}{3 \pi} \frac{1}{\sqrt{-2E_{\inf}}-{\alpha}{\rm arctanh}\left({\alpha}/\sqrt{-2 E_{\inf}}\right)}.
\end{eqnarray}
One can easily read from this result that $E_{\rm inf}$ becomes more negative with
increasing SOC strength.

\begin{figure*}[tbp]
\includegraphics[width=5.5cm]{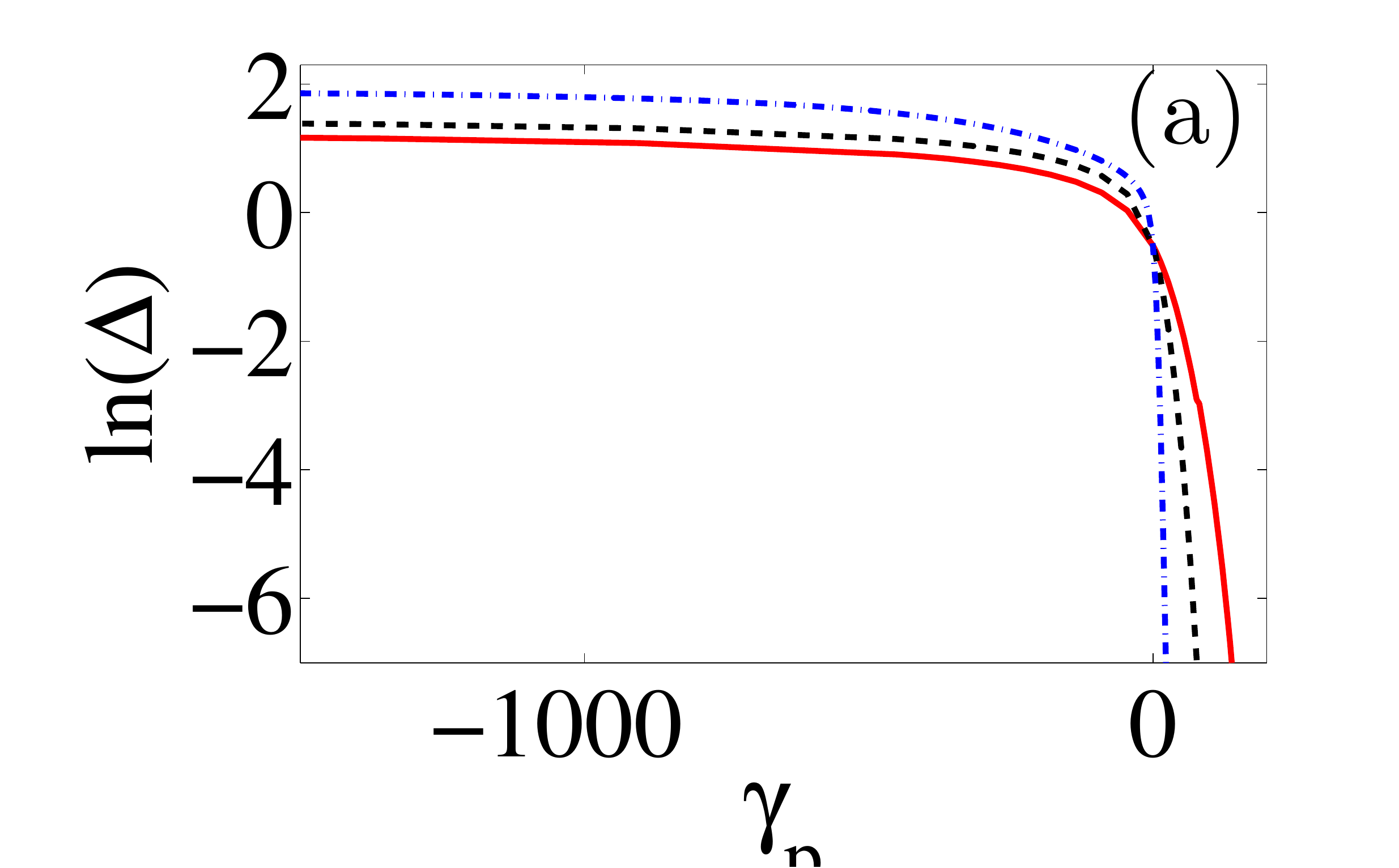}
\includegraphics[width=5.5cm]{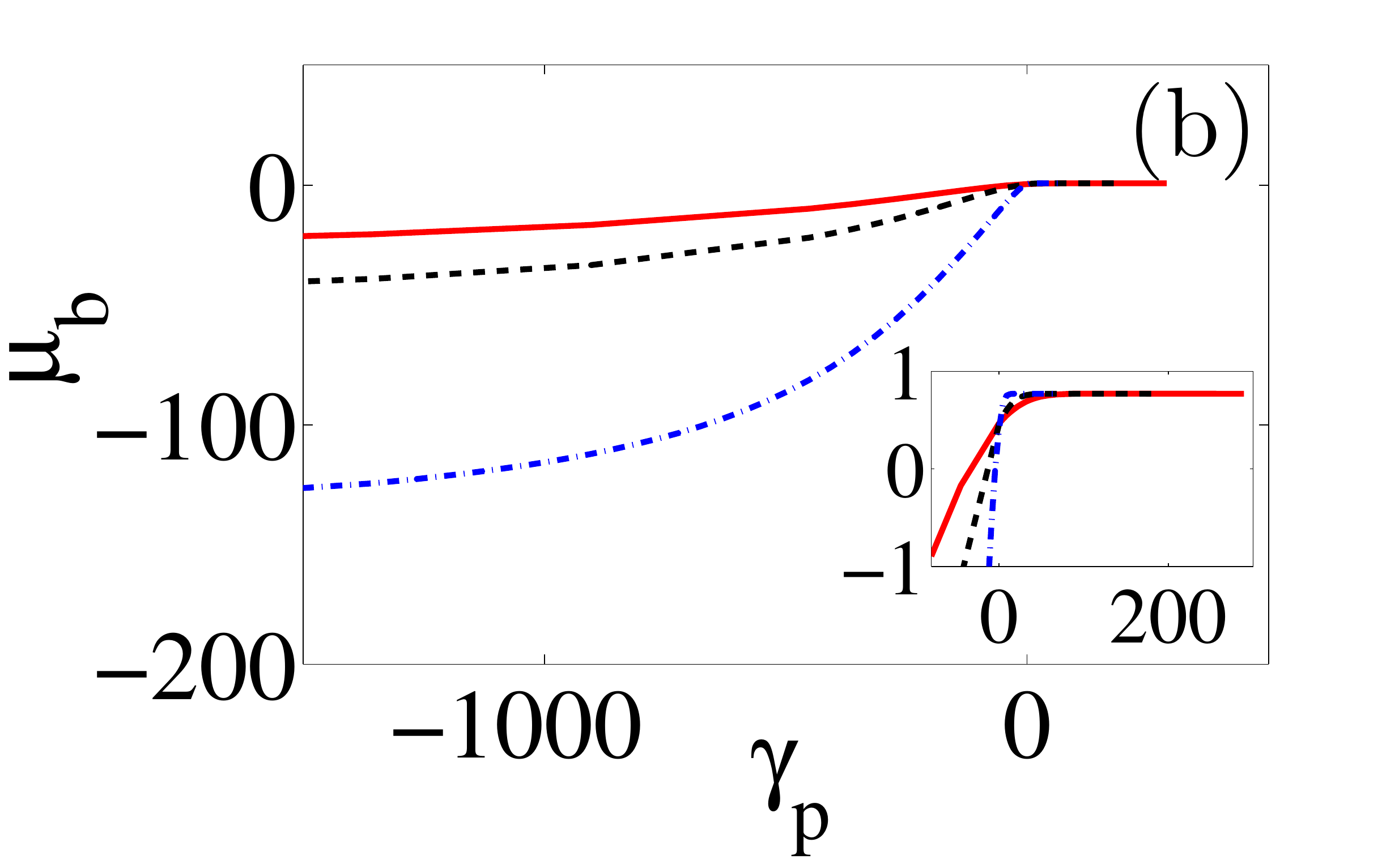}
\includegraphics[width=5.5cm]{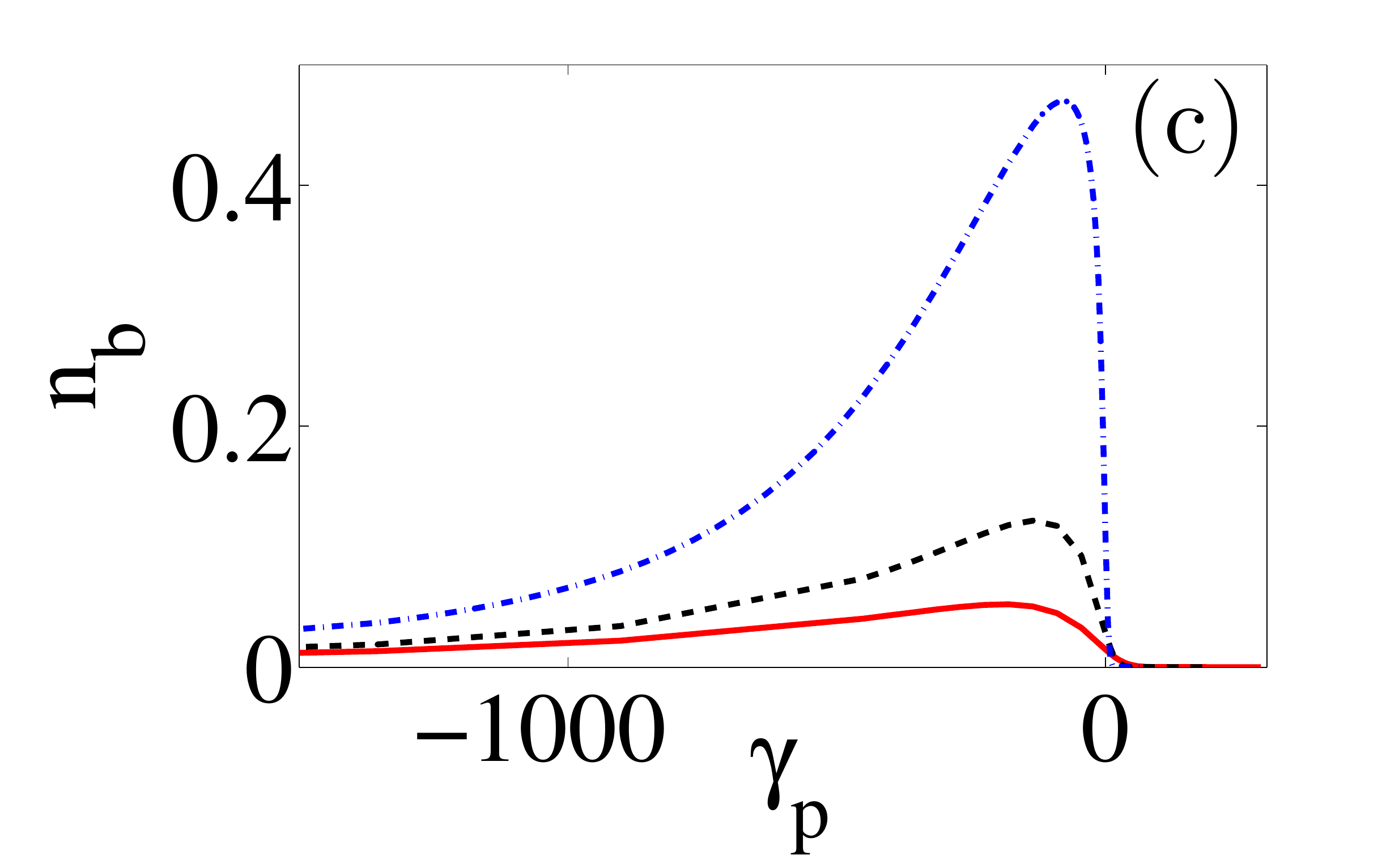}
\includegraphics[width=5.5cm]{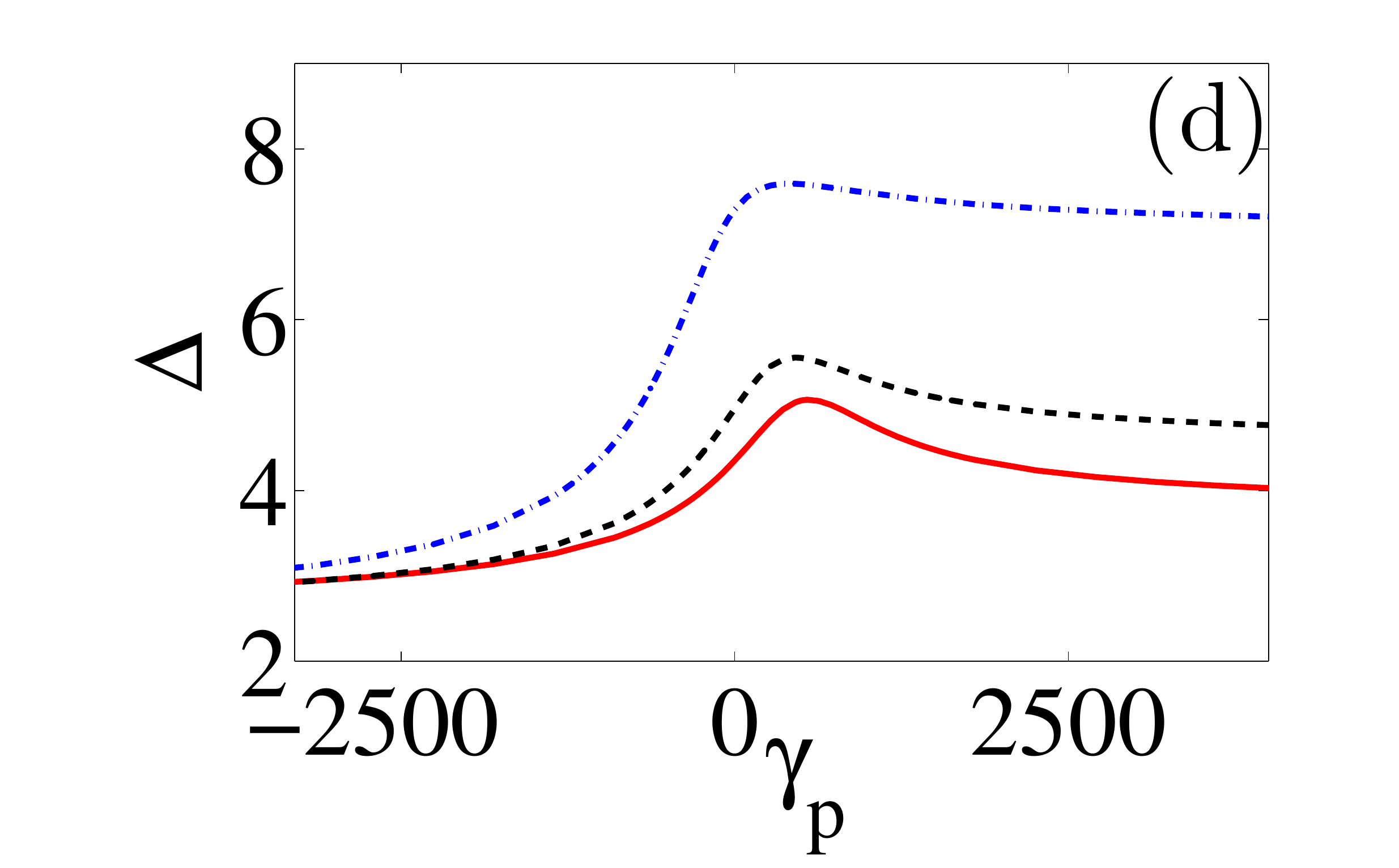}
\includegraphics[width=5.5cm]{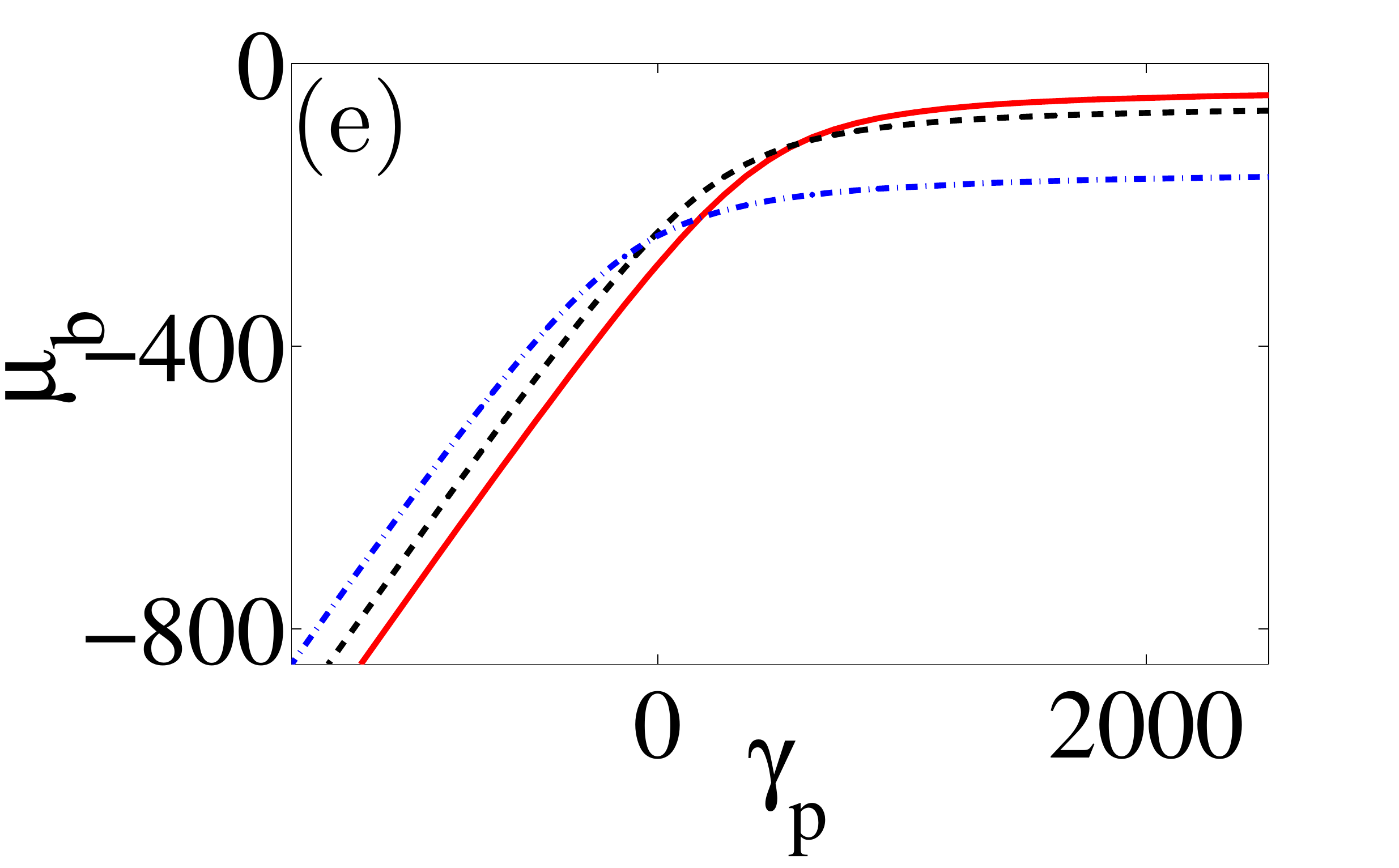}
\includegraphics[width=5.5cm]{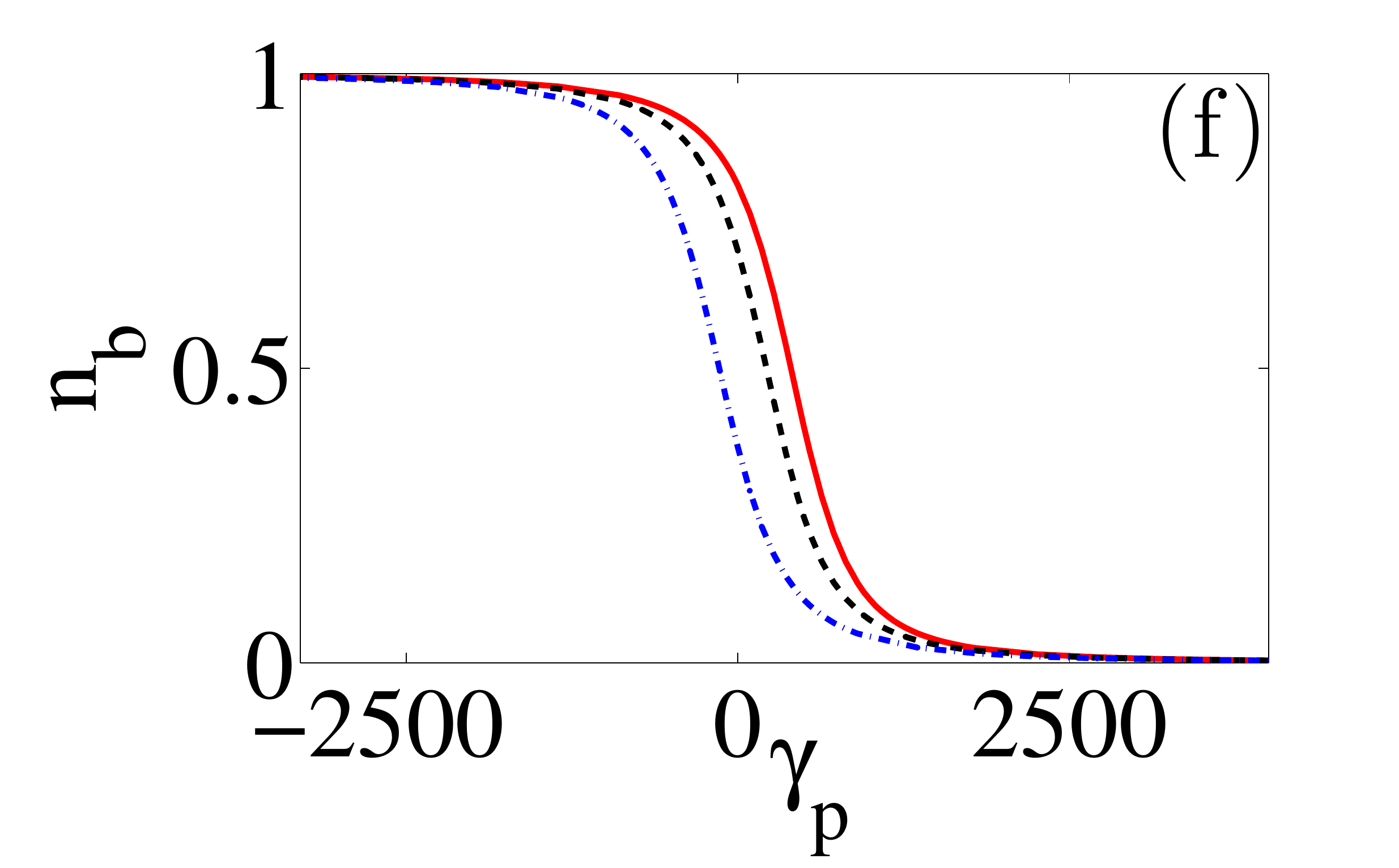}
\caption{(Color online)  The superfluid order parameter $\Delta$, the shifted chemical potential $\mu_b \equiv \mu + \alpha^2/4$,
and the molecular fraction $n_b$ associated with (a-c) the upper branch solution, and (d-f) the lower branch solution
for the case of a positive background scattering length. For all panels, dimensionless parameters are chose as
$U_p=0.17, g_p=7, \alpha=1$ (solid, red), $U_p=0.14, g_p=5, \alpha=1$ (dashed, black),
and $U_p=0.1, g_p=2.5, \alpha=1$ (dash-dotted, blue).}\label{ABGZUG}
\end{figure*}

\section{Manby-body pairing states}

In this section, we characterize the many-body properties of the system at zero temperature. Following the standard BCS mean-field theory, the effective Hamiltonian can be written in a matrix form in the pseudo-spin basis $\{a_{\mathbf{k},\uparrow},a_{-\mathbf{k},\uparrow}^{\dag},a_{\mathbf{k},\downarrow} a_{-\mathbf{k},
\downarrow}^{\dag}\}^{T}$:
\begin{eqnarray}
H_{\text{eff}}-{\mu} N
&=&\frac{1}{2}\sum_{\mathbf{k}} \left[\begin{array}{cccc}\lambda_{\mathbf{k}}&{\Delta}&0&\kappa_{\mathbf{k}}^{-}\\ {\Delta}&-\lambda_{\mathbf{k}}&\kappa_{\mathbf{k}}^{-}&0\\ 0&\kappa_{\mathbf{k}}^{+}&-\lambda_{\mathbf{k}}&-{\Delta}\\ \kappa_{\mathbf{k}}^{+}&0&-{\Delta}&\lambda_{\mathbf{k}}\end{array}\right] \nonumber\\
&&\hspace{-2.5cm}
+\sum_{\mathbf{k}}\left(\epsilon_{\mathbf{k}}+{U}f \right)+({\gamma}-2{\mu})|\psi_m|^2- {U}(|p|^2+f^2).\label{H0}
\end{eqnarray}
Here, $\lambda_{\mathbf{k}}=\epsilon_{\mathbf{k}}-{\mu}+Uf, \kappa_{\mathbf{k}}^{\pm}={\alpha}(k_x\pm ik_y)$, the order parameter ${\Delta}={g}\psi_m+{U}p$, and the mean-field parameters are defined as:
\begin{align}
\psi_m=&<b_0>\nonumber\\
p=&\sum_{\mathbf{k}}<a_{-\mathbf{k},\downarrow}a_{\mathbf{k},\uparrow}>\nonumber\\
f=&\sum_{\mathbf{k}}<a^{\dag}_{\mathbf{k},\uparrow}a_{\mathbf{k},\uparrow}> =\sum_{\mathbf{k}}<a^{\dag}_{\mathbf{k},\downarrow}a_{\mathbf{k},\downarrow}>.\label{FieldParameter}
\end{align}
Notice that the dimensionless total particle number $N=1$ in the unit system defined in Sec. II.

By diagonalizing the effective Hamiltonian Eq. (\ref{H0}) and by imposing the renormalization condition with physical parameters,
we obtain the expression for the ground state thermodynamic potential at zero temperature
\begin{align}
\Omega=&\sum_{\mathbf{k}} \left[ \epsilon_{\mathbf{k}}-\frac{1}{2}(E_{\mathbf{k},+}+E_{\mathbf{k},-})+{U}f \right]
\nonumber\\
&+ ({\gamma}-2{\mu})|\psi_m|^2 -{U}(|p|^2+f^2),
\end{align}
where the quasi-particle dispersion $E_{\mathbf{k},\pm}=\sqrt{A^2_{\mathbf{k},\pm}+{\Delta}^2}$,
and $A_{\mathbf{k},\pm}=\epsilon_{\mathbf{k}}-{\mu}+{U} f\pm {\alpha} k_{\perp}$ with $k_\perp = \sqrt{k_x^2 + k_z^2}$
is defined to simplify notation.

From the extrema conditions $\partial\Omega/\partial f=0$, $\partial\Omega/\partial p=0$, $\partial\Omega/\partial\psi_m=0$, and the number equation
$N=-\partial\Omega/\partial{\mu}$, we have a set of self-consistent equations \cite{huhuitwo-c}:
\begin{align}
\psi_m=&-\frac{{g}_p p}{ ({\gamma}_p-2{\mu}) }, \nonumber\\
2f=&\sum_{\mathbf{k}}\left(1-\frac{A_+}{2E_+}-\frac{A_-}{2E_-}\right), \nonumber\\
1= &\left( {U}_p-\frac{{g}^2_p}{{\gamma}_p-2{\mu}}\right) \sum_{\mathbf{k}}
\left(\frac{1}{2\epsilon_{\mathbf{k}}}-\frac{1}{4E_{\mathbf{k},+}} -\frac{1}{4E_{\mathbf{k},-}}\right), \nonumber\\
1=&2f+2|{\Delta}|^2   \left[{g}_p-\frac{({\gamma}_p-2{\mu}){U}_p}{{g}_p} \right]^{-2},\label{EQUF}
\end{align}
from which the ground state parameters can be determined. Note that for the parameter regime discussed in this work, we find the influence on the chemical potential induced by the Hartree term $f$ remains negligible. Thus, the quasi-particle dispersion can be approximated as
\begin{equation}
E_{\mathbf{k},\pm} \approx \sqrt{  (\epsilon_{\mathbf{k}}-{\mu} \pm {\alpha} k_{\perp})^2  +{\Delta}^2}.
\end{equation}
Note that the self-consistent Eqs. (\ref{EQUF}) can be reduced to the more familiar forms of the gap and number equations under SOC in
a single-channel model by setting ${g}_p=0$ \cite{wy2d}.
We also define the closed channel fraction as
\begin{equation}
n_b=2 |{\Delta}|^2 \left[({g}_p-\frac{({\gamma}_p-2{\mu}){U}_p}{{g}_p})^2 \right]^{-1},
\end{equation}
which will be used to describe the properties of the underlying system.

\begin{figure*}[tbp]
\includegraphics[width=5.5cm]{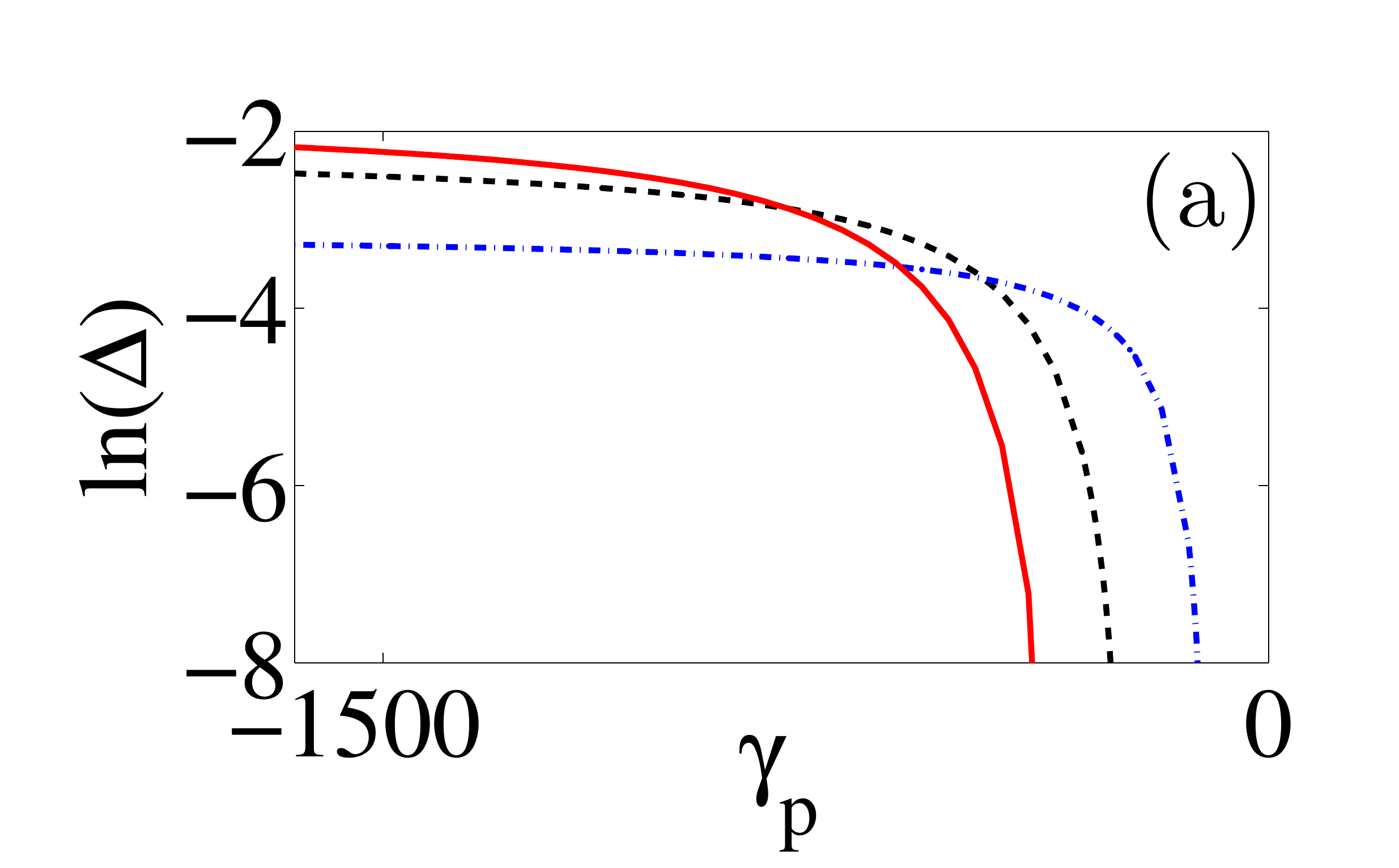}
\includegraphics[width=5.5cm]{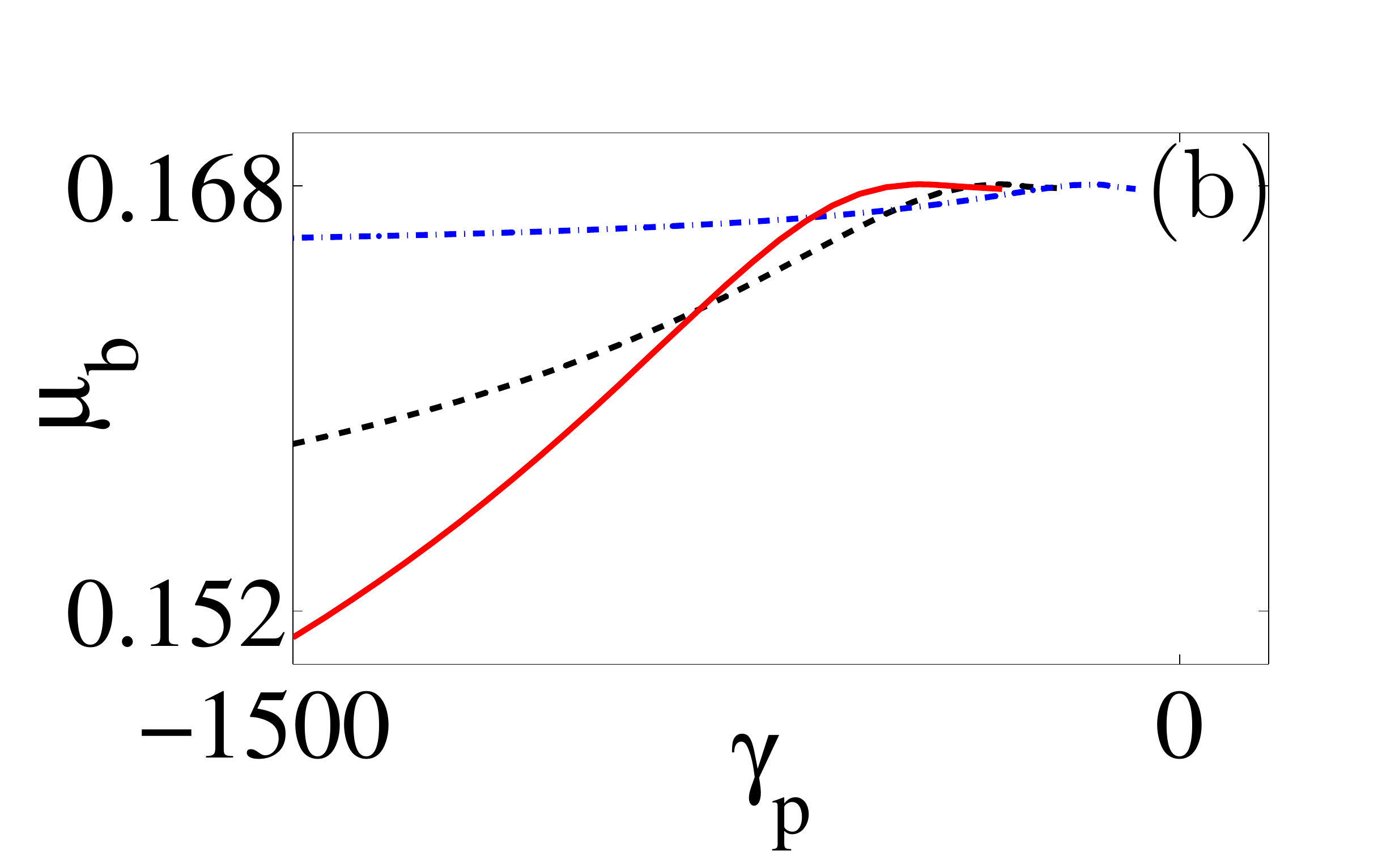}
\includegraphics[width=5.5cm]{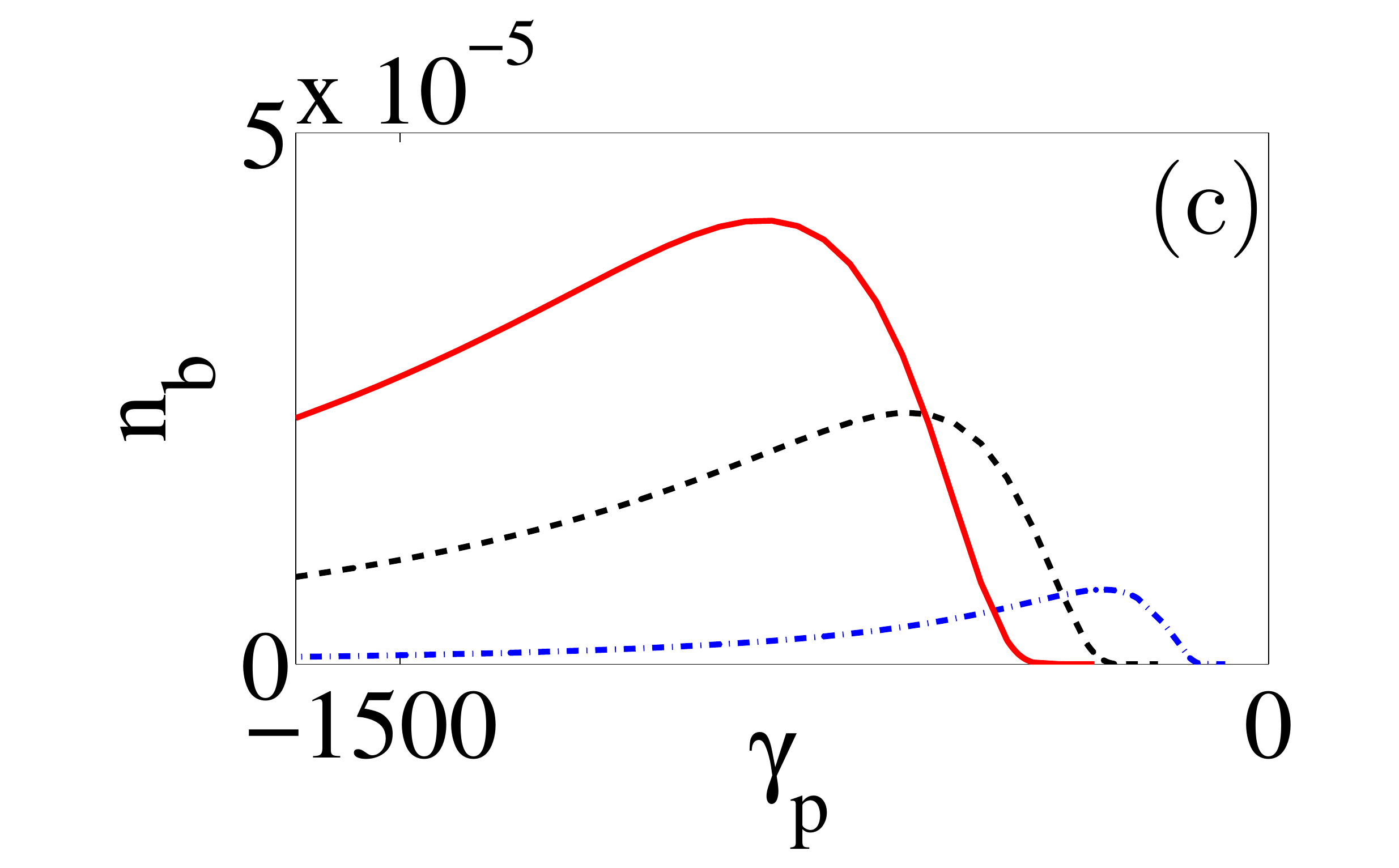}
\includegraphics[width=5.5cm]{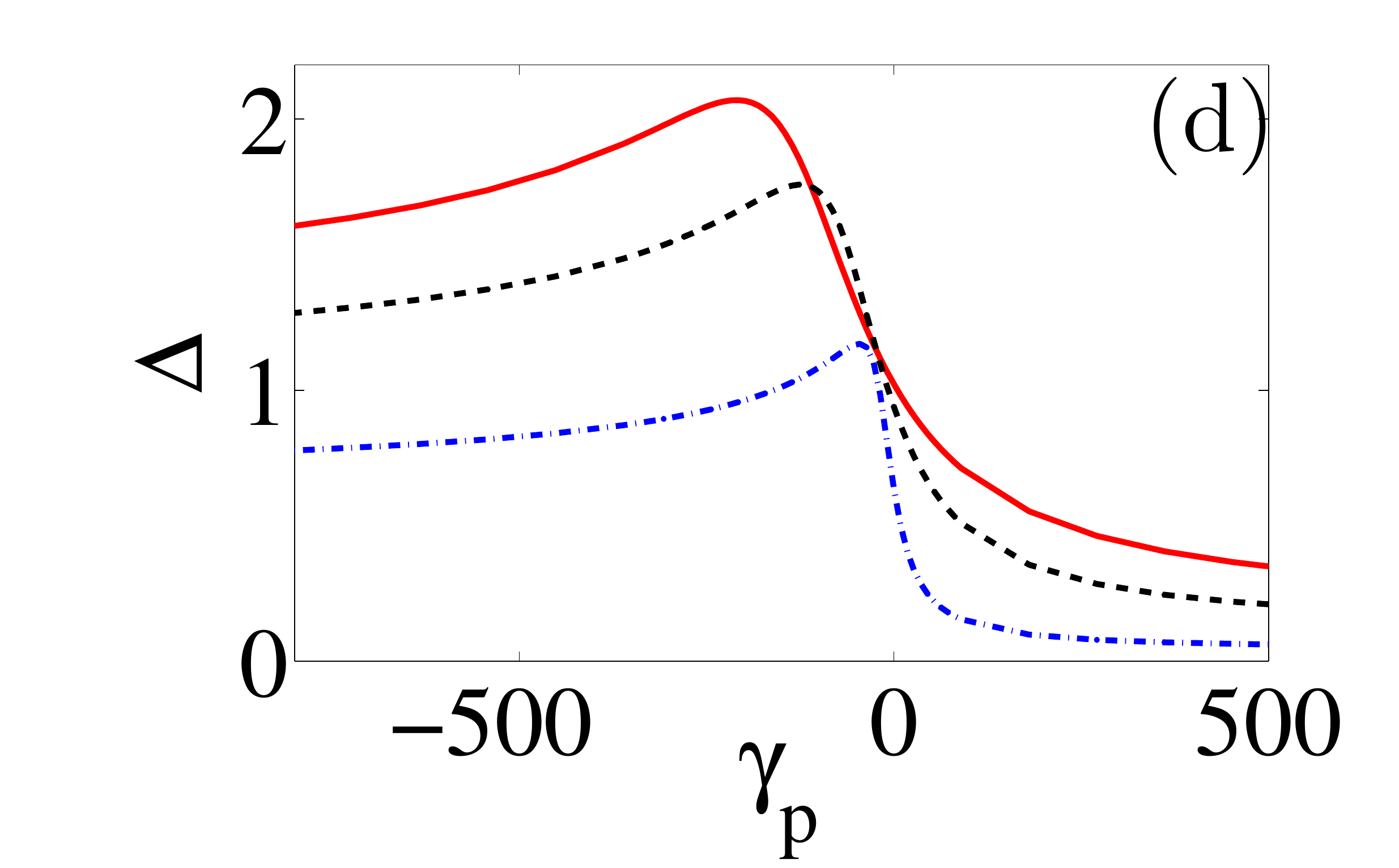}
\includegraphics[width=5.5cm]{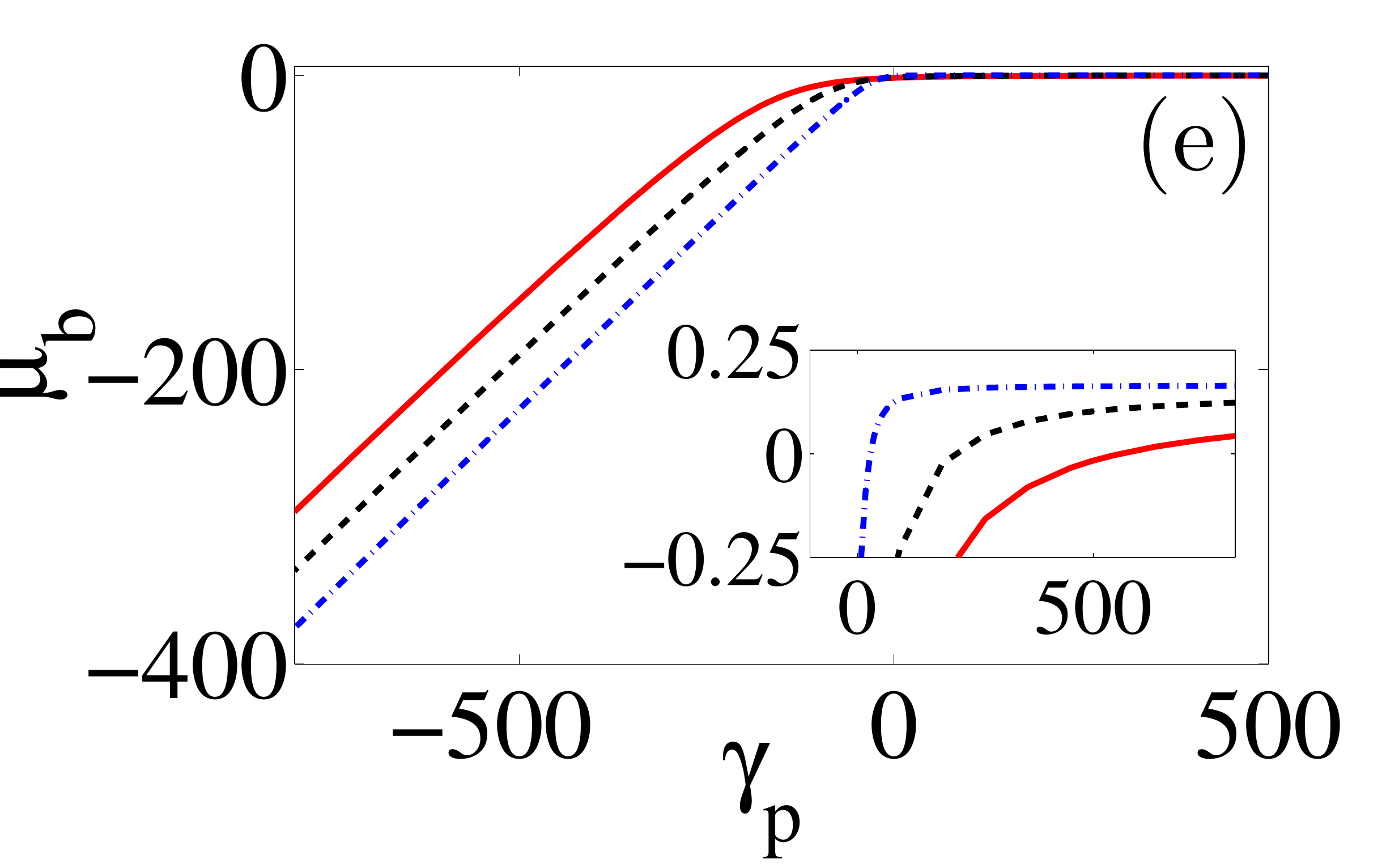}
\includegraphics[width=5.5cm]{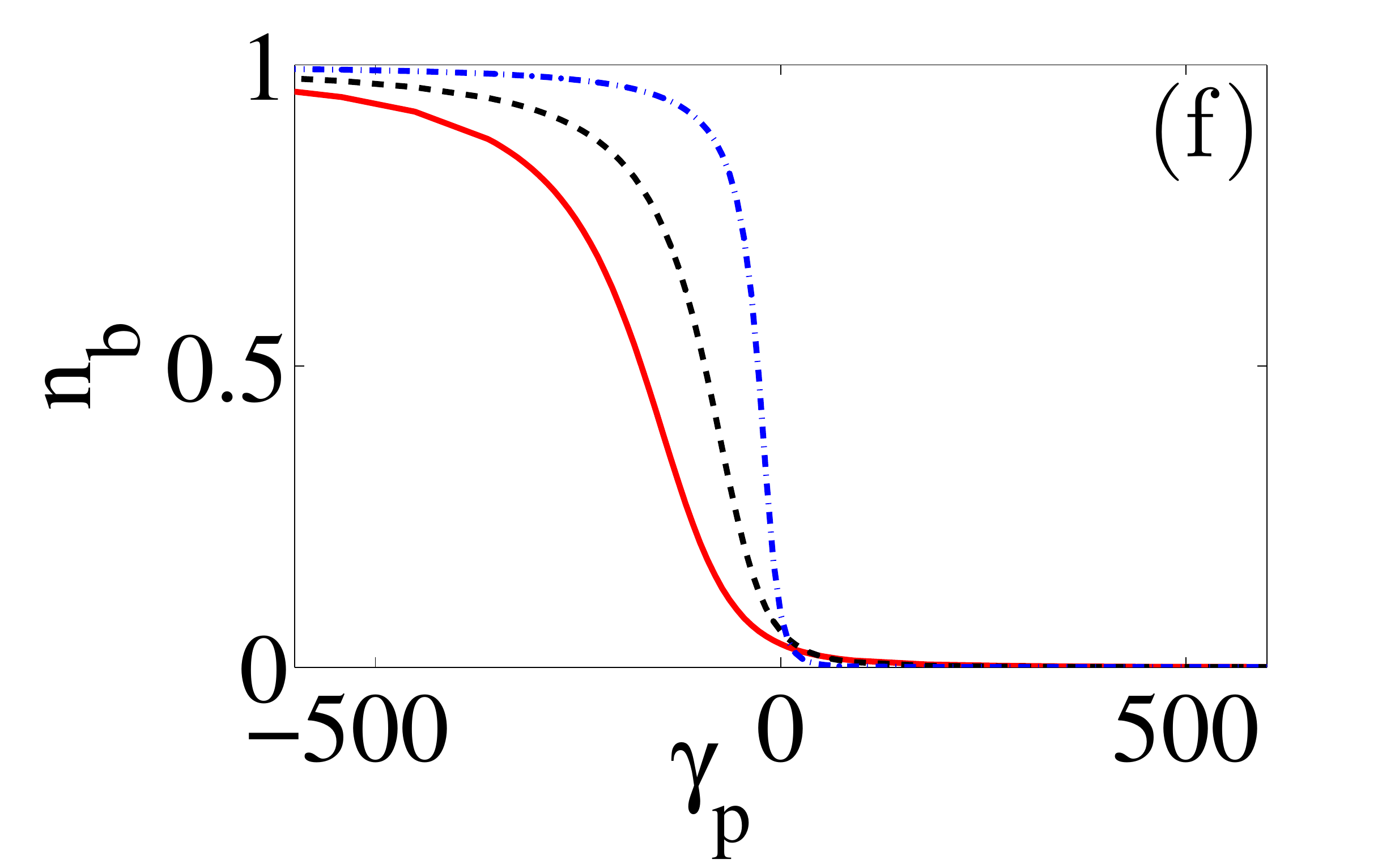}
\caption{(Color online) The superfluid order parameter $\Delta$, the shifted chemical potential $\mu_b \equiv \mu + \alpha^2/4$,
and the molecular fraction $n_b$ associated with (a-c) the upper branch solution, and (d-f) the lower branch solution for
the case of a negative background scattering length. Dimensionless parameters used in this figure are
$U_p=-0.17, g_p=7, \alpha=5$ (solid, red), $U_p=-0.14, g_p=5, \alpha=5$ (dashed, black), and $U_p=-0.1, g_p=2.5, \alpha=5$
(dash-dotted, blue). }\label{ABGFUG}
\end{figure*}

\subsection{Positive background scattering length}

For the case of a positive background scattering length $a_{\rm bg} > 0$, there exists a weakly-bound state in the open channel away from the Fesbach resonance. As the magnetic field is tuned close to the resonance point, the coupling between the bound states in the open and the closed channels gives rise to the two branches of many-body solution in the absence of SOC. Previous studies have shown the existence of a quantum phase transition in the upper branch where the many-body ground state changes from a superfluid state to a normal state \cite{WY_D}. The qualitative picture remains valid in the presence of SOC, where quantitative modifications can be induced by the SOC. In Fig. \ref{ABGZUG}, we map out various mean-field quantities as functions of the detuning for several scattering parameters.

In Fig. \ref{ABGZUG}(a-c) we show the results of order parameter, the chemical potential, and the molecular fraction as functions of detuning for the upper branch, which corresponds to the weakly-bound state in the two-body case.
Here, the chemical potential is plotted after subtracting the single-particle threshold with SOC, i.e., $\mu_b \equiv \mu + \alpha^2/4$.
An interesting feature here is the existence of a quantum phase transition, whose location can be identified as the detuning where the order parameter approaches zero [see Fig. \ref{ABGZUG} (a)]. As the resonance width narrows, the location of the phase transition point moves towards the BEC-side of the Feshbach resonance. Importantly, with appropriate resonance width and SOC strength, the location of the quantum phase transition point may cross the Feshbach resonance and reach the BEC side, as we will show in Sec. V. We also note that the shifted chemical potential $\mu_b$ in the upper branch crosses zero on the BEC side of the resonance, demonstrating the existence of a BCS-BEC crossover [see the inset of Fig. \ref{ABGZUG}(b)]. We show in Fig. \ref{ABGZUG}(d-f) the properties of the lower branch, which corresponds to the deeply-bound state in the two-body case. It is apparent that the shifted chemical potential stays negative with an order parameter approaching finite values in both the weak and strong coupling limit. Physically, the solution in the lower branch corresponds to a condensate of tightly bound molecules, which become Rashbons in the large SOC limit~\cite{soc3}.

\subsection{Negative scattering length}

We now turn to the case with negative background scattering length $a_{\rm bg}<0$. In the absence of SOC, there is only one branch of many-body solution, which features a BCS-BEC crossover as the interaction strength is tuned \cite{WY_D}. When SOC is turned on, however, this picture is drastically modified. Similar to the two-body case, a new branch (upper branch) of many-body solution emerges. Interestingly, a quantum phase transition can also be identified in this upper branch.

In Fig. \ref{ABGFUG} (a-c) we show the results of the order parameter, the shifted chemical potential, and the molecular fraction as functions of detuning for the
upper branch. Here, the location of the quantum phase transition is pushed towards the BEC-limit with increasing SOC strength or Feshbach resonance width,
and the shifted chemical potential $\mu_b$ remains positive for arbitrary detuning.
In Fig. \ref{ABGFUG} (d-f), we show the same quantities for the lower branch. Notice that there is no quantum phase transition in this branch, as the order parameter is always finite. For small SOC, the shifted chemical potential is positive in the BCS-limit, indicating the existence of a Fermi surface. Hence, the system in the lower branch undergoes a BCS-BEC crossover as the interaction becomes stronger or as the SOC strength increases.


\section{Quantum Phase Transition}

In previous sections, we see that the quantum phase transition in the upper branch is intimately connected with the SOC. For positive background scattering length, SOC can modify the location of the phase transition point, while for negative background scattering length, SOC can induce a new quantum phase transition. In this section, we discuss in detail the dependence of the phase transition point on various parameters.

The condition for the onset of this quantum phase transition can in fact be obtained analytically by examining the gap and the number equations (\ref{EQUF}). At the critical detuning where the quantum phase transition occurs, we have $\Delta=0$. For the upper branch, regardless of the sign of the background scattering length, this is only possible when the right-hand-side of the gap equation also tends to infinity as the quantum critical point is approached. Therefore, at the critical point, the denominator of the left-hand-side of the gap equation must vanish, leading to
\begin{align}
U_p=\frac{g^2_p}{\gamma_p^c-2\mu}.\label{EQg}
\end{align}
Here, $\gamma_p^c$ is the critical detuning of the quantum phase transition point in the upper branch.
Similarly, the number equation at the critical point takes the form
\begin{align}
1=\sum_{\mathbf{k}}\left(1-\frac{A'_{k,+}}{2|A'_{k,+}|}-\frac{A'_{k,-}}{2|A'_{k,-}|}\right),\label{EQn}
\end{align}
where $A'_{\mathbf{k},\pm}=\epsilon_{\mathbf{k}}-{\mu}\pm \alpha k_{\perp}$. From these equations, we see that SOC only affects the quantum critical point through the chemical potential $\mu$.

\begin{figure}[tbp]
\includegraphics[width=8.4cm]{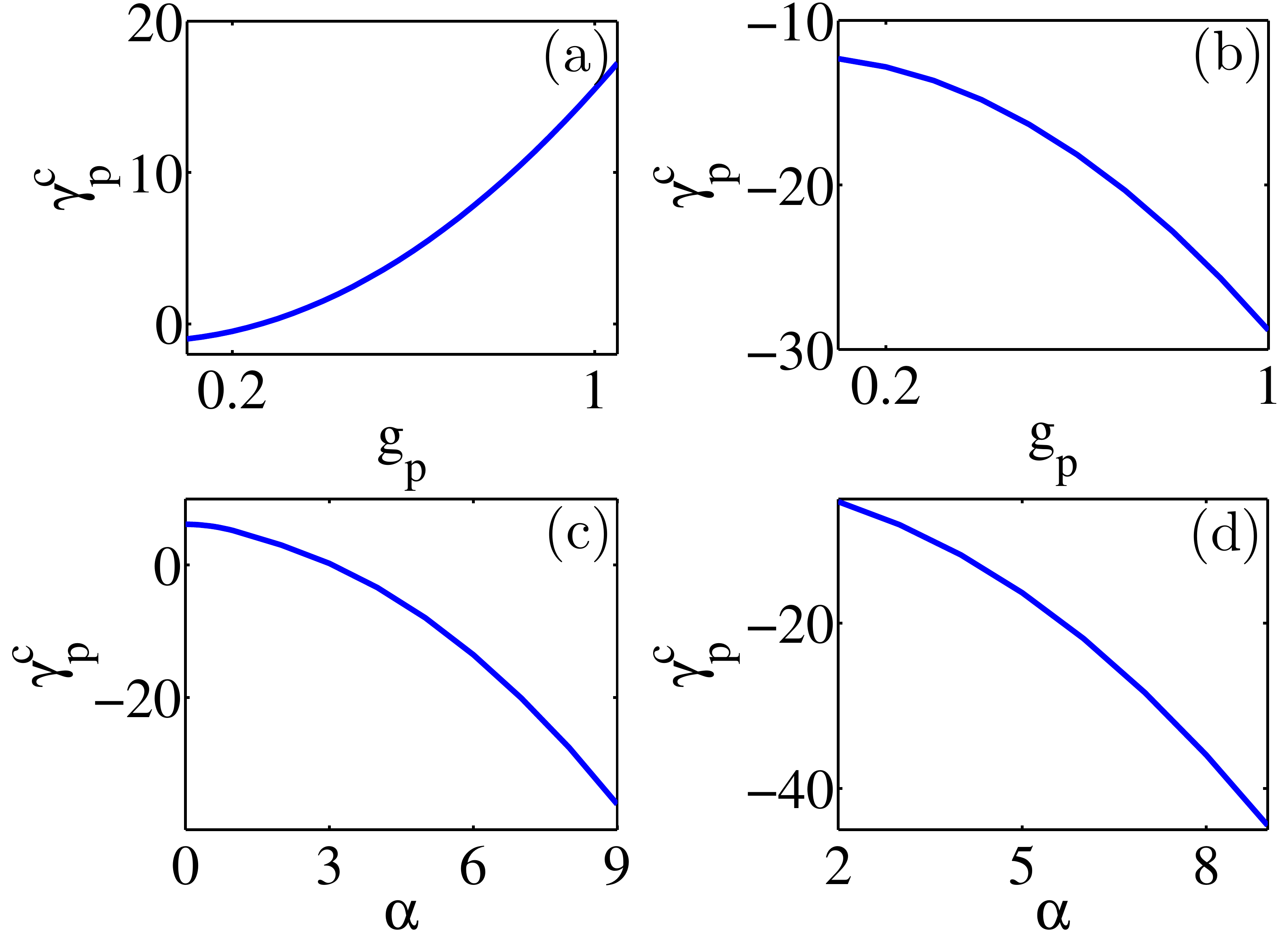}
\caption{Critical detuning $\gamma_p^c$ for the quantum phase transition in the upper branch as functions of $g_p$ and $\alpha$.
Dimensionless parameters used in these plots are (a) $\alpha=2, U_p=0.06$; (b) $\alpha=5, U_p=-0.06$;
(c) $g_p=0.5, U_p=0.06$; and (d)  $g_p=0.5, U_p=-0.06$.
} \label{QPT}
\end{figure}

We show in Fig. \ref{QPT} the phase transition point in the upper branch as functions of the scattering parameter $g_p$ and the SOC strength.
For a fixed background interaction rate $U_p$, the width of the Feshbach resonance typically narrows with decreasing $g_p$.
For a positive background scattering length, the critical detuning can be made to cross the resonance point when either the resonance width or the SOC strength is tuned [see Fig. \ref{QPT}(a) and \ref{QPT}(c)].
For systems with negative background scattering length, however, the critical point is always lying on the BEC side of the Feshbach resonance
in any realistic situation.
By choosing a narrow resonance with small SOC strength, this quantum phase transition can be tuned closer to the resonance points, as
can be seen in Fig. \ref{QPT}(b) and \ref{QPT}(d). This observation suggests that for the experimental observation of this quantum phase transition,
a system with narrow Feshbach resonance under moderate SOC strength should be preferred.

\section{SUMMARY}
We have studied a spin-orbit coupled ultracold Fermi gas near a Feshbach resonance by using a two-channel model. We find that under a finite SOC and with a finite background scattering length, there are in general two branches of solution for both the two-body and the many-body ground states. This is in contrast to the conventional BCS-BEC crossover picture, where the background scattering length is typically neglected; and is different from the case without SOC, where the upper-branch solution only exists for a positive background scattering length. As a result, for a negative background scattering length, the bound state in the open channel is purely SOC induced. These lead to the interesting situation that a quantum phase transition exists in the upper branch of the many-body solution. The location of the quantum phase transition can be tuned by the SOC strength, or by choosing Feshbach resonances with different resonance widths. In particular, the critical point of the quantum phase transition can be tuned close to or across the resonance point, where the Fermi gas is most stable against three-body losses. It is therefore hopeful that such phase transitions can be observed in experiments.

\acknowledgments
This work is supported by NFRP (2011CB921200, 2011CBA00200), NKBRP (2013CB922000), NNSF (60921091), NSFC (11105134, 11274009, 11374283), SRFDP (20113402120022), the Fundamental Research Funds for the Central Universities (WK2470000006), and the Research Funds of Renmin University of China (10XNL016, 13XNH123).

\end{document}